\documentclass[a4paper,12pt]{article}
\usepackage[american]{babel}
\usepackage[latin1]{inputenc}
\usepackage[T1]{fontenc}
\usepackage{graphicx}
\usepackage{epsfig}
\usepackage{amsmath}

\title{Brane-world cosmology and varying $G$}
\author{Leonardo Amarilla$^{1,2}$~ and Héctor Vucetich$^{2}$}

\begin{document}
\maketitle

\begin{center}
\begin{small}
$^{1}$Facultad de Ciencias Exactas y Naturales, Universidad de Buenos Aires - \emph{Pabellón 1, Ciudad Universitaria (1428). Buenos Aires, Argentina}\\
$^{2}$Facultad de Ciencias Astronómicas y Geofísicas, Universidad Nacional de La \nolinebreak Plata - \emph{Paseo del Bosque S/N (1900). La Plata, Argentina}
\end{small}
\end{center}

  We consider a brane-world cosmological model coupled to a bulk scalar
  field. Since the brane tension turns out to be proportional
  to Newton coupling $G$, in such a model a time variation of $G$ naturally
  occurs. By resorting to available bounds on the variation of $G$, the
  parameters of the model are constrained. The constraints coming
  from nucleosynthesis and CMB result to be the severest ones.\\



\section{Introduction}

The first ideas of a varying Newton's coupling $G$ come from 1937, when P. A. M. Dirac
introduced his famous Large Numbers Hypothesis
\cite{diracLNH1,diracLNH2}. In the 60s, in an attempt to reconciliate
Mach Principle with General Relativity (GR), Brans and Dicke
developed their well known scalar-tensor theory of gravity
\cite{BD}. Following Jordan's ideas, Brans and Dicke generalized GR including a
varying $G$, whose dynamics was governed by a scalar field.
See \cite{uzan} for a detailed review on varying fundamental constants.

In addition to the effects of introducing a dynamical coupling constant,
we know that gravitational interaction is also sensitive to the existence
of extra dimensions, which could manifest themselves at short distances.
In this paper, we will be concerned with models that incorporate both a
varying $G$ and a higher dimensional set up.

Although the idea of extra dimensions is not new either \cite{KK1,KK2}, the advent of modern
(string) theories has brought to the fore the higher dimensional scenarios more recently.
One of the string inspired models that have attracted much attention in the last decade is
the Randall-Sundrum model (RS), which consists of an effective
brane-world embedded in an orbifold of AdS$_{5}$ space \cite{RSalternative}.
The property of the RS-like models that is interesting for our purpose is the relation
between the tension of the brane, and the Newton constant
of the 4-dimensional effective theory. Models with non-constant brane tension thus lead to a
time-varying $G$, as we will discuss below.

The idea of this paper is to confront particular brane-world models with
constraints coming form observational cosmological data. The particular
model we will consider here is that of \cite{review}, which is
motivated by supergravity in singular spaces. We will consider this
model as a working example to show how observational data could be
used to constrain parameters of this type of scenarios.

The observational data we will use to constrain the model are of rather
different types. For instance, we have data coming from planetary/geological scale:
Observations due to space missions to Mercury, Mars, Venus and the Moon
in the 70s, determined that if $G$ varies in time, its variation is
less than $10^{-11}$ per year. On the other hand, in the late 70s, many works appeared
relating the relative variation of $G$ with planetary radius \cite{egyed,mcelh}.
McElhinny \emph{et al.} studied the evolution of the Earth's radius and
extended their study to the Moon, Mars and Mercury, and constrained $\Delta
G/G$ in specific moments close to Solar System formation.

At cosmological scale, a variation of $G$ leads to modifications in
the Friedmann equation. The direct consequences of these variations
are changes in the cooling rate of the universe and in the computed
primordial abundances of He and Li. In \cite{accetta}, Accetta \emph{et al.}
used this relation between varying $G$ and light
elements abundances to give a bound to the relative variation of $G$.
This variation (its absolute value, in fact) turns to be less than 40\% since
Big-Bang nucleosynthesis (BBN). CMB anisotropies are also sensitive to a varying $G$.  Chan \emph{et
al.} concluded in \cite{chan-chu} that the relative variation of $G$
since recombination is less than 10\%.

In this work, we explore a five-dimensional gravitational model,
\emph{alla} RS, with a scalar field in the bulk that modificates the
brane tension, which induces a variation in $G$. The variations of $G$
predicted by the model, depending on two parameters, will be then compared
with the observational data mentioned above. That is, the aim is to
constrain the possible values of these parameters, using experimental bounds.

This work is organized as follows: In Section 2 we discuss
the Randall-Sundrum-like  model coupled to a scalar field, which
induces variation of effective Newton constant in four dimensions.
In Section 3 we survey bounds on the variations of $G$ and the observations.
In Section 4, we combine observational data of Section 3 with the predictions
of the model, and use this to constrain the parameters.
In particular, we give bounds on the 5D-Planck mass, supersymmetry  breaking scale,
and the cosmological constant in the bulk. In Section 5, we draw some conclusions.

\section{A brane-world scenario}

\subsection{Field equations}

The RS-like scenarios propose that our universe is a 3-brane embedded in
a curved asymptotically hyperbolic 5-dimensional bulk, or an orbifold of it.
One can also include matter in the brane as well as in the bulk \cite{review};
here we consider a scalar field $\phi$.
The brane is located at the origin of the fifth dimension, which we denote $x_{5}$. This
dimension has \textbf{$Z_{2}$} symmetry in our case.

Fields of the Standard Model live on the brane, while gravitational interaction
(and the scalar field) is free to propagate in the bulk. Bulk action is given by

\begin{equation}\label{eq:Sbulk}
S_{\text{bulk}}=\frac{1}{2\kappa_{5}^{2}}\int{{d^{5}x}\sqrt{-g_{(5)}}}
\bigg(R-\frac{3}{4}\Big((\partial{\phi})^{2}+U(\phi)\Big)\bigg),
\end{equation}
where $R$ is the curvature scalar associated to the 5-dimensional metric
$g^{(5)}_{AB}$; $U(\phi)$ is the bulk potential, which depends on
the scalar field $\phi$, and \\ $\kappa_{5}^{2}=1/M_{5}^{3}$, being
$M_{5}$ the Planck mass in 5D.

The action of the brane depends on its tension $U_{B}(\phi)$ (brane
potential). In our case, it is a function of the scalar $\phi$ and of the
confined matter; namely
\begin{equation}\label{eq:Sbrana}
S_{\text{brane}}=\int{d^{4}x}\sqrt{-g_{(4)}}
\Big(-\frac{3}{2\kappa_{5}^{2}}U_{\text{B}}(\phi(x_{5}=0))+
L_{\text{matter}}\Big),
\end{equation}
with
$g_{(4)}^{\mu\nu}=\delta_{M}^{\mu}\delta_{N}^{\nu}g_{(5)}^{MN}|_{x_{5}=0}$.
In this paper, Latin indices in capital letters go from 0 to 5
(excluding 4), Greek indices go from 0 to 3, and Latin indices in
regular letters, from 1 to 3.

The matter content of the 5D space is characterized by the
energy-momentum tensor, which can be derived from the total action and
has the bulk and brane contributions; namely
\begin{equation}\label{eq:tensorem}
T_{AB}=T^{\mathrm{\text{bulk}}}_{AB}+T^{\mathrm{\text{brane}}}_{AB},
\end{equation}
with
\begin{equation}\label{eq:TAB}
T^{\mathrm{\text{bulk}}\; A}_{\phantom{\mathrm{\emph{bulk}}\;
A}B} = \frac{3}{4}\left(\partial^{A}\phi \partial_{B}\phi - \frac{1}{2}g^{\mathrm{(5)}\; A}_{\phantom{\mathrm{(5)}\; A}B}
(\partial\phi)^{2}\right)-\frac{3}{8}g^{\mathrm{(5)}\; A}_{\phantom{\mathrm{(5)}\; A}B}U(\phi),
\end{equation}

\begin{equation}\label{eq:Tbrana}
T^{\mathrm{\text{brane}}\; A}_{\phantom{\mathrm{\text{brane}}\;
A}B} = \left(-\frac{3}{2}g^{\mathrm{(5)}\; A}_{\phantom{\mathrm{(5)}\; A}B}
U_{\text{B}}(\phi)+\tau^{\mathrm{\text{matter}}~A}_{\phantom{\mathrm{\text{matter}}\;
A}B}\right) \delta(x_{5}),
\end{equation}
and
\begin{equation}\label{eq:taubrana}
\tau^{\mathrm{\text{matter}}\;A}_{\phantom{\mathrm{\text{matter}}\;A}B} =
diag(-\rho_{\text{m}},p_{\text{m}},p_{\text{m}},p_{\text{m}},0).
\end{equation}
Tensor $\tau^{\text{matter}}$ is related to the ordinary matter on the brane.

The energy density $\rho_{\text{m}}$ and the pressure $p_{\text{m}}$ are independent
of the position in the brane, so one recovers an homogeneous
cosmology in four dimensions. The equation of state that relates these
quantities is taken to be $p_{\text{m}}=\omega_{\text{m}}\rho_{\text{m}}$.

Einstein equations reads
\begin{equation}\label{eq:Ein}
  G_{AB}\equiv
  R_{AB} - \frac{1}{2}R~g^{(5)}_{AB} =
  \kappa_{5}^{2}\left(T^{\mathrm{\text{bulk}}}_{AB} + T^{\mathrm{\text{brane}}}_{AB}
  \right)=\kappa_{5}^{2}T_{AB}.
\end{equation}

Let us propose the following ansatz for the metric:
\begin{equation}\label{eq:metricagral}
ds^{2}=-A^{2}(t,x_{5})dt^{2}+
B^{2}(t,x_{5})dx_{i}dx^{i}+C^{2}(t,x_{5})dx_{5}^{2}.
\end{equation}

We are interested in cosmological scale solutions, so we
assume an isotropic and homogeneous metric in the three spatial
coordinates. That is
\begin{equation}\label{eq:metrica}
ds^{2} = a^{2}(t,x_{5}) b^{2}(x_{5}) (-dt^{2} + dx_{5}^{2}) +
a^{2}(t,x_{5})\Omega_{ij}dx^{i}dx^{j},
\end{equation}
where $\Omega_{ij}$ is the metric of a 3-dimensional space with
constant curvature:
\begin{equation}\label{eq:omega}
\Omega_{ij}=\delta_{ij}\Big(1+\frac{K}{4}x^{l}x^{m}\delta_{lm}\Big)^{-2},
\end{equation}
where the values $K=0,1,-1$ correspond to a (spatially) flat, closed or open
universe, respectively.  Since observational evidences are consistent
with a spatially flat Universe \cite{wmap}, then we assume $K=0$.

It is important to note that in \eqref{eq:metrica} we chose a conformal gauge for the
($0-5$) part of the metric. In this gauge, the brane is placed in a
fixed position, $x_{5}=0$, \emph{i.e.} the fixed point of the $Z_{2}$
symmetry in the fifth dimension.
Function $b$ only depends on the spatial coordinate $x_{5}$.

To derive the brane dynamics, one must verify that, although the
equations of motion must be restricted to it, these equations have to be
satisfied in the bulk as well.  The brane "proper time" is
\begin{equation}\label{eq:dtau}
d\tau=ab|_{x_{5}=0}dt,
\end{equation}
and the differential of the normal vector to its surface is given by
\begin{equation}\label{eq:dy}
dy=ab|_{x_{5}=0}dx_{5}.
\end{equation}
From now on, we write $\dot{f}=\frac{df}{d\tau}$, $f'=\frac{df}{dy}$.

The Israel-Darm\^{o}ise junction conditions describe how a brane with
a given energy-momentum tensor can be embedded in a higher dimensional
space-time. These equations yield

\begin{equation}\label{eq:isr1}
\frac{a'}{a}|_{y=0}=-\frac{1}{6}\kappa_{5}^{2}\rho,
\end{equation}
\begin{equation}\label{eq:isr2}
\frac{b'}{b}|_{y=0}=\frac{1}{2}\kappa_{5}^{2}(\rho+p),
\end{equation}
where equations $\rho$ is the energy density, and $p$ is the pressure on the brane.

On the other hand, the boundary condition for the scalar $\phi$ is
\cite{csaki-erlich-grojean-hollowood}
\begin{equation}\label{eq:contphi}
\phi'|_{y=0}=\frac{\partial U_{\text{B}}}{\partial\phi}|_{y=0}.
\end{equation}

The total energy density and pressure on the brane can be written
as a sum of two contributions: a term related to confined matter,
and a second one, related to the tension, which in this case
depends on the scalar field. Thus, we have
\begin{equation}\label{eq:rhop}
\rho=\rho_{\text{m}}+\frac{3}{2\kappa_{5}^{2}}U_{\text{B}},
~p=p_{\text{m}}-\frac{3}{2\kappa_{5}^{2}}U_{\text{B}}.
\end{equation}

In what follows, all quantities will be evaluated on the
brane, \emph{i.e.} at $y=0$. Restricting the ($0-5$) component of
the Einstein equations to the brane,
and using boundary conditions \eqref{eq:isr1} and \eqref{eq:isr2}
we obtain the energy conservation equation
\begin{equation}\label{eq:cons}
\dot{\rho}=-3H(\rho+p)-2T^{0}_{\phantom{0}5},
\end{equation}
where $H\equiv\frac{\dot{a}}{a}|_{y=0}$ is the Hubble parameter on the brane.

Using the explicit form for $\rho$ and $p$ from \eqref{eq:rhop}, total
energy density conservation law transforms into a conservation law for
the energy of the brane; that is,
\begin{equation}\label{eq:rhom}
\dot{\rho_{\text{m}}}=-3H(\rho_{\text{m}}+p_{\text{m}}).
\end{equation}
Time variation of the scalar field energy density $3U_{\text{B}}/2$ cancels
the term involving $T^{0}_{\phantom{0}5}$, since the latter can be
written as
$T^{0}_{\phantom{0}5}=-\frac{3}{4}\phi'\dot{\phi}=-\frac{3}{4}\dot{U}_{\text{B}}$.

The solution for the brane energy conservation is
\begin{equation}\label{eq:rhodea}
\rho_{\text{m}}=\rho_{0}a^{-3(1+\omega_{\text{m}})},
\end{equation}
as in standard cosmology.

On the other hand the Friedmann equation on the brane is a consequence of the $(5-5)$ component
of Einstein equations.
For a brane containing matter coupled to a scalar field  $\phi$,
Friedmann equation reads
\begin{equation}\label{eq:fr4}
H^{2} = \frac{\kappa_{5}^{4}}{36} \rho_{\text{m}}^{2} + \frac{\kappa_{5}^{2}} {12} U_{\text{B}}
\rho_{\text{m}} - \frac{1}{16a^{4}} \int d\tau
\frac{da^{4}}{d\tau} (\dot{\phi}^{2}-2V) - \frac{\kappa_{5}^{2}}{12a^{4}}\int
d\tau a^{4}\rho_{\text{m}} \frac{dU_{\text{B}}}{d\tau}+ \frac{A}{a^{4}},
\end{equation}
with
\begin{equation}\label{eq:defV}
V=\frac{1}{2}\Big(U_{\text{B}}^{2}-\big(\frac{\partial U_{\text{B}}}{\partial
\phi}\big)^{2}+U \Big).
\end{equation}

The set of equations is completed by the Klein-Gordon equation
for the scalar field \cite{brax-vdbruck-davis,mennim-batye}; namely
\begin{equation}\label{eq:KG}
\ddot{\phi}+4H\dot{\phi}+\frac{1}{2}\big(\frac{1}{3}-\omega_{\text{m}}\big)\rho_{\text{m}}\frac{\partial
U_{\text{B}}}{\partial \phi}\kappa_{5}^{2}=-\frac{\partial V}{\partial\phi}
+ \Delta \Phi,
\end{equation}
where
\begin{equation}\label{eq:deltPhi}
\Delta \Phi = \frac{\partial^{2}\phi}{\partial
y^{2}}|_{y=0}-\frac{\partial U_{\text{B}}}{\partial
\phi}|_{y=0}\frac{\partial^{2} U_{\text{B}}}{\partial \phi^{2}}|_{y=0}.
\end{equation}

Following \cite{brax-vdbruck-davis} and \cite{mennim-batye}, we consider
\begin{equation}\label{eq:Phinulo}
\Delta\Phi=0.
\end{equation}

Einstein equations have been used to write \eqref{eq:KG} in this form
(see \cite{brax-vdbruck-davis}).

\subsection{Physical considerations}

Friedmann equation \eqref{eq:fr4} is not conventional. In contrast to
the standard one, \eqref{eq:fr4} presents terms that depend on the field $\phi$,
a quadratic term in the energy density on the brane
(present also in absence of the scalar), and an
additional term that goes like $a^{-4}$.

By the time of primordial nucleosynthesis, corrections coming from brane models,
including the term proportional to the square of the energy density in
Friedmann equation, must be negligible. Otherwise, the rate of
expansion would be modified and the computation of light elements
abundances would be inconsistent with observations.  In this
non-conventional scenario, the freezing temperature of proton to
neutron ratio $T_{C}$ would be of the order of $(2-3)$ MeV, while in
standard cosmology it is $T_{C} \sim (0.7-0.8)$ MeV, consistent with He
abundance.  The difference between both temperatures is a direct
consequence of the fact that Hubble parameter is linear with $T^{4}$,
and not with $T^{2}$, generating a slower cooling of the Universe
\cite{binetruy-deffayet-langlois-nonconv}.  However, corrections might
be important during the inflationary period.

Let us be reminded of the fact that in standard cosmology Friedmann equation is
\begin{equation}\label{eq:fristand}
H^{2}_{\text{stand}}=\frac{8 \pi G}{3}\rho_{\text{m}}+ \frac{\Lambda_{4}}{3},
\end{equation}
where $\Lambda_{4}$ is the cosmological constant in four dimensions.
Then, the quadratic term in $\rho_{\text{m}}$ in \eqref{eq:fr4}
can be identified with the first term in \eqref{eq:fristand}; that is
\begin{equation}\label{eq:ident}
\frac{U_{\text{B}}(\phi)}{12}\kappa_{5}^{2}= \frac{8\pi G}{3}.
\end{equation}

It is clear that in our model, Newton constant in 4D varies as it depends on
$\phi$; \emph{i.e.} it is possible to find time variation of $G$.

\subsubsection{Bulk and brane potentials}

We consider a functional form for the potential $U(\phi)$ coming from
the supergravity models in singular spaces studied in
\cite{csaki-erlich-grojean-hollowood}.  Following these results, one finds

\begin{equation}\label{eq:U}
U=\left(\frac{\partial{W}}{\partial{\phi}}\right)^{2} - W^{2},
\end{equation}
where $W(\phi)$ is the so called superpotential.

We study the case in which the superpotential in an exponential
function of the field
\begin{equation}\label{eq:W1}
W(\phi) = 4k e^{\alpha\phi},
\end{equation}
where $[k^{-1}]=L$ and
$\alpha$ is a real number.

The brane potential is defined through the superpotential by
\begin{equation}\label{eq:UB}
U_{\text{B}}=TW,
\end{equation}
where $T$ is a real number related to the scale of supersymmetry breaking
\cite{brax-davis}.

Having the functional relation between $U_{B}$ and the
scalar field, given by \eqref{eq:W1} and \eqref{eq:UB},
one can find $G(\phi)$ using \eqref{eq:ident}; thus,
\begin{equation}\label{eq:Gdephi}
G(\phi)=\frac{k}{8 \pi} \kappa_{5}^{2} T e^{\alpha \phi}.
\end{equation}

The expectation value of $\phi$ today is assumed to be zero by convention as a
boundary condition. Then, Newton "constant" would be given by
\begin{equation}\label{eq:Gdephi}
G_{\text{today}}(\phi)=\frac{k}{8 \pi} \kappa_{5}^{2} T.
\end{equation}

\subsubsection{Working hypothesis}

The form of the Friedmann equation with all the new
contributions is quite abstruse. Then, in order to solve the model,
some approximations and assumptions have to be taken into account.
We discuss these below.

First, the term proportional to $a^{-4}$, can be considered as
a correction to the radiation density.
This term is usually referred as dark radiation.
Here, we assume $A=0$ in \eqref{eq:fr4}.
We also consider a low energy regime, \emph{i.e.} we neglect the term proportional to
$\rho_{\text{m}}^{2}$ in \eqref{eq:fr4}.
It is possible to do this under the condition $\rho_{\text{m}}\ll \rho_{crit}$, with
\begin{equation}\label{eq:denscrit}
\rho_{\text{crit}}=\frac{3U_{\text{B}}}{\kappa_{5}^{2}}=\frac{12}{\kappa_{5}^{2}}k T
\sim 4.6 × 10^{33}\frac{g}{cm^{3}},
\end{equation}
where \eqref{eq:W1}, \eqref{eq:UB}, and bounds on $k$, $\kappa_{5}^{2}$ and
$T$ consistently found \emph{a posteriori} in Sections 3 and 4, have been used.  In the studied period
(between BBN and today) the density remains below
this critical density.
We also assume that the time evolution of the scalar field $\phi$ in
the brane proper time $\tau$ is much slower than the one of the scale
factor $a$. It is possible to extract $\dot{\phi}^{2}$ from the first
integral in \eqref{eq:fr4} in this adiabatic regime.

A non-dissipative approximation of the potential will be also considered.
The brane potential $U_{\text{B}}$ is basically Newton constant on the brane, up to
multiplicative constants. Following the adiabatic approximation,
it is reasonable to suppose that the contribution correspondent to this
term might be negligible.
The term $dU_{\text{B}}/d\tau$, as well as the other terms of order
$\dot{\phi}$ and $\dot{a}$, contribute to higher order estimation
of $G(\phi)$ than the one we study here.

Finally, and consequently with the assumptions above, the square of the time derivative
of $\phi$, \emph{i.e.} the kinetic energy of the field, is lower than other terms in
Friedmann equation.
It is possible to make a simply calculation to constrain the current value
of the time derivative of the scalar field:
Following the approximations, Friedmann equation today, divided by $H_{0}^{2}$ is
\begin{equation}\label{eq:frie}
1=\Omega_{\text{M}} + \Omega_{\text{R}} + \Omega_{\Lambda} +
\Omega_{\partial\phi},
\end{equation}
where $\Omega_{\partial\phi}=\dot{\phi}^{2}(\tau_{0})/16$.

Recent data \cite{wmap} implies that the sum of the first three contributions is close to $0.996$,
which fixes a limit to the absolute current value of the time derivative of $\phi$; namely
\begin{equation}\label{eq:cotafi0}
|\frac{\dot{\phi}}{H_{0}} (\tau_{0})|< 0.22.
\end{equation}
This heuristic argument supports our approximation hypothesis.

In addition, the exponential dependences appearing in Friedmann and
Klein-Gordon equations will be approximated to $1$ since $\phi$ is small.

With all the approximations described above, Friedmann equation takes the form
\begin{equation}\label{eq:fr5}
H^{2}=T\frac{k}{3}\kappa_{5}^{2}\rho_{\text{m}}+\frac{\Lambda_{\text{eff}}}{3},
\end{equation}
with
\begin{equation}\label{eq:lamdaeff}
\frac{\Lambda_{\text{eff}}}{3}:=k^{2}(T^{2}-1)(1-\alpha^{2}).
\end{equation}

On the other hand, eq.\eqref{eq:rhodea} says that, for radiation,
\begin{equation}\label{eq:rhorad}
\rho_{\text{R}}= \rho_{0\text{R}}H_{0}^{4}a^{-4},
\end{equation}
and for non-relativistic matter,
\begin{equation}\label{eq:rhomat}
\rho_{\text{M}}= \rho_{0\text{M}}H_{0}^{4}a^{-3},
\end{equation}
with $\rho_{0\text{M},\text{R}}$ dimensionless constants.

Thus, Friedmann equation can be rewritten as follows:
\begin{equation}\label{eq:frMRL}
H^{2} = T \frac{k}{3} \kappa_{5}^{2} \left(\frac{\rho_{0\text{M}}}{a^{3}} +
\frac{\rho_{0\text{R}}}{a^{4}}\right) H_{0}^{4} + \frac{\Lambda_{\text{eff}}}{3}.
\end{equation}
After making the identifications
\begin{equation}\label{eq:identrhoomega}
T\frac{k}{3}\kappa_{5}^{2}\rho_{0\text{M},\text{R}}H_{0}^{2}= \Omega_{\text{M},\text{R}},
\end{equation}
\begin{equation}\label{eq:identlam}
\frac{\Lambda_{\text{eff}}}{3H_{0}^{2}}= \Omega_{\Lambda},
\end{equation}
we obtain the familiar form for the equation
\begin{equation}\label{eq:frstan}
\frac{H^{2}}{H_{0}^{2}} = \frac{\Omega_{\text{M}}}{a^{3}} + \frac{\Omega_{\text{R}}}{a^{4}} +
\Omega_{\Lambda}.
\end{equation}
Then, the equation for $\phi$ is

\begin{equation}\label{eq:KG2}
\ddot{\phi}+4H\dot{\phi}+2\big(\frac{1}{3}-\omega_{\text{m}}\big)\rho_{\text{m}}\kappa_{5}^{2}\alpha
k T=-16 \alpha \frac{\Lambda_{\text{eff}}}{3}.
\end{equation}

The system of equations above can now be solved in two different epochs: one
dominated by radiation and matter, and the other governed by matter
and cosmological constant.

\subsection{Solution to the Field equations}

\subsubsection{Radiation and matter dominated epoch}

In this epoch the term proportional to
$\Lambda_{\text{eff}}$ can be neglected. Then, the Fridmann equation takes the form\footnote{Subindex $1$ refers to the radiation-matter regime.}
\begin{equation}\label{eq:frMRstan1}
\frac{H_{1}^{2}}{H_{0}^{2}} = \frac{\Omega_{\text{M}}}{a_{1}^{3}} +
\frac{\Omega_{\text{R}}}{a_{1}^{4}},
\end{equation}
whose solution is
\begin{equation}\label{eq:a1eta}
a_{1}(\eta) = \frac{\Omega_{\text{M}}}{4}(\eta-\eta_{0})^{2} - \frac{\Omega_{\text{R}}}{\Omega_{\text{M}}},
\end{equation}
where the integration has been performed in the conformal time $\eta$, being
$d\tau=\frac{a_{1}(\eta)}{H_{0}} d\eta$.

The equation of state for radiation is $p_{\text{rad}}=\rho_{\text{rad}}/3$,
while $p_{\text{mat}}=0$ is the equation for non-relativistic dust. Then, KG equation reads:
\begin{equation}\label{eq:KGMR}
\ddot{\phi_{1}}+4H_{1}\dot{\phi_{1}}+2\alpha
H_{0}^{2}\frac{\Omega_{\text{M}}}{a_{1}^{3}}=-16 \alpha H_{0}^{2}
\Omega_{\Lambda},
\end{equation}
where we used eqs. \eqref{eq:identrhoomega} y  \eqref{eq:identlam}.
The solution for the differential equation is (see Appendix)
\begin{equation}\label{eq:fi1}
\phi_{1}(a_{1})=B-\frac{A}{2\sqrt{\Omega_{\text{R}}}a_{1}^{2}}-\frac{2\Omega_{\text{M}}\alpha}{3\Omega_{\text{R}}}a_{1},
\end{equation}
where it has been taken into account that scale factor remains small
during this regime.

\subsubsection{Matter and cosmological constant dominated epoch}

During this regime, the term proportional to the inverse of the fourth
power of the scale factor $a$ is negligible when compared to the other
two terms. Then, the Friedmann equation reads\footnote{Subindex $2$
refers to the matter-$\Lambda$ regime.}
\begin{equation}\label{eq:frMLstan}
\frac{H_{2}^{2}}{H_{0}^{2}}=\frac{\Omega_{\text{M}}}{a_{2}^{3}}+\Omega_{\Lambda}.
\end{equation}

Integrating, one obtains the scale factor during this epoch, namely
\begin{equation}\label{eq:a2tau}
a_{2}(\tau) = a_{0} \sinh^{\frac{2}{3}}
\left(\frac{\sqrt{\Omega_{\Lambda}}}{2}H_{0}(\tau-C_{0})\right).
\end{equation}
In this regime, the Klein-Gordon equation is
\begin{equation}\label{eq:KGML}
\ddot{\phi}+4H_{2}\dot{\phi}+2\alpha
H_{0}^{2}\frac{\Omega_{\text{M}}}{a_{2}^{3}}=-16 \alpha H_{0}^{2}
\Omega_{\Lambda}.
\end{equation}

In order to solve this equation it is convenient to consider
two cases: $a_{2}\ll 1$, that occurs for times close to union time
($\tau \sim \tau_{\text{u}}$); and $a_{2} \sim 1 $, it is, times near today
($\tau \sim \tau_{0}$):
\begin{itemize}
\item $a_{2} \ll 1$:

\begin{equation}\label{eq:fi2}
\phi_{2}(a_{2})=D-\frac{2C}{5\sqrt{\Omega_{\text{M}}}a_{2}^{\frac{5}{2}}}-\frac{4}{5}\alpha
\ln (a_{2}).
\end{equation}

\item $a_{2} \sim 1$:
\begin{equation}\label{eq:fi3}
\phi_{3}(a_{2}) = G - \frac{F}{3\sqrt{\Omega_{\Lambda}+\Omega_{\text{M}}}a_{2}^{3}} -
\frac{\alpha}{5(\Omega_{\Lambda}+\Omega_{\text{M}})}(8\Omega_{\Lambda}+\Omega_{\text{M}})a_{2}^{2},
\end{equation}
\end{itemize}
where $G$ and $F$ are integration constants.
Details of this calculation can be found in the Appendix.

\subsubsection{Boundary conditions for the scale factor}

To find the integration constants in \eqref{eq:a1eta} and \eqref{eq:a2tau},
one must define the boundary conditions. We take the convention
\begin{equation}\label{eq:ahoy}
a_{2}(\tau_{0})=1,
\end{equation}
where $\tau_{0}$ is the current value of $\tau$.
We also have the matching conditions
\begin{equation}\label{eq:Hhoy}
H_{2}(\tau_{0})=H_{0},
\end{equation}
and
\begin{equation}\label{eq:Hu}
H_{1}(\tau_{\text{u}})=H_{2}(\tau_{\text{u}}),
\end{equation}
where $\tau_{\text{u}}$ is the junction time between both regimes.
In addition, we have
\begin{equation}\label{eq:au}
a_{1}(\tau_{\text{u}})=a_{2}(\tau_{\text{u}})=a_{\text{u}}.
\end{equation}

Constants $a_{0}$ and $C_{0}$ can be obtained from
\eqref{eq:ahoy} and \eqref{eq:Hhoy}. Consequently, we have
\begin{equation}\label{eq:a2cc}
a_{2}(\tau) = \left(\frac{1}{\Omega_{\Lambda}}-1\right)^{\frac{1}{3}}
\sinh^{\frac{2}{3}}\left(\frac{3\sqrt{\Omega_{\Lambda}}}{2}H_{0} \left(\tau-\tau_{0} + \frac{2}{3H_{0}\sqrt{\Omega_{\Lambda}}}\sinh^{-1}
\left(\sqrt{\frac{\Omega_{\Lambda}}{1-\Omega_{\Lambda}}}\right)\right)\right).
\end{equation}
The constant of integration must be chosen to satisfy the condition\\
$a_{1}(\tau=0)=0$. Then
\begin{equation}\label{taudea2}
\tau(a_{1}) = \frac{2}{3H_{0}\Omega_{\text{M}}^{2}}(\Omega_{\text{M}}a_{1} -2\Omega_{\text{R}})
\sqrt{\Omega_{\text{M}}a_{1} + \Omega_{\text{R}}}+ \frac{4\Omega_{\text{R}}^{\frac{3}{2}}}{3H_{0}\Omega_{\text{M}}^{2}}.
\end{equation}

\begin{figure}[!h]
\begin{center}
\includegraphics[width=1.1\linewidth]{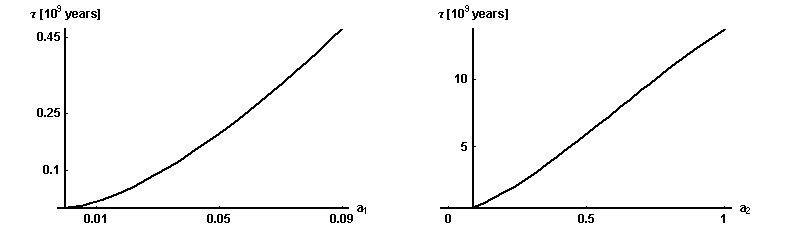}
\caption{Graphics for $\tau$ vs $a_{1}$ and $\tau$ vs $a_{2}$}\label{fig:taudea1}
\end{center}
\end{figure}

\subsubsection{Boundary conditions for the scalar field}

Since the scalar field must be smooth, $\phi_{1}$ (and its derivative) must be equal to $\phi_{2}$ (resp. to
its derivative) in the junction time $\tau_{\text{u}}$, when\\
$a=a_{\text{u}}=\Omega_{\text{R}}/\Omega_{\Lambda}$. That is,
\begin{equation}\label{eq:ficcau}
\phi_{1}(a_{\text{u}})=\phi_{2}(a_{\text{u}}),
\end{equation}
\begin{equation}\label{eq:dficcau}
\frac{d\phi_{1}}{da_{1}}(a_{\text{u}})=\frac{d\phi_{2}}{da_{2}}(a_{\text{u}}).
\end{equation}

The field $\phi_{2}$ must be equal to $\phi_{3}$ (as well
as their derivatives) in an intermediate time $\tau_{\text{I}}$
between $\tau_{\text{u}}$ and $\tau_{0}$. We take $\tau_{\text{I}}$
as the time equidistant to $\tau_{\text{u}}$ and $\tau_{0}$. Thus,
\begin{equation}\label{eq:ficcai}
\phi_{2}(a_{\text{I}})=\phi_{3}(a_{\text{I}}),
\end{equation}
\begin{equation}\label{eq:dficcai}
\frac{d\phi_{2}}{da_{2}}(a_{\text{I}})=\frac{d\phi_{3}}{da_{2}}(a_{\text{I}}).
\end{equation}

The last boundary conditions we need are the values of the field and
its derivative today, \emph{i.e.} at $\tau_{0}$ ($a(\tau_{0})=a_{0}=1$):
\begin{equation}\label{eq:ficca0}
\phi_{3}(a_{0})=0,
\end{equation}
\begin{equation}\label{eq:dficca0}
\frac{d\phi_{3}}{da_{2}}(a_{0})=\frac{\dot{\phi_{0}}}{H_{0}}:=p_{0}.
\end{equation}

Using \eqref{eq:ficcau}-\eqref{eq:dficca0} and the
values for $H_{0}$, $\Omega_{\text{M}}$, $\Omega_{\text{R}}$ and $\Omega_{\Lambda}$ listed
in Subsection 3 of the Appendix, one obtains the constants of integration
$A,B,C,D,F$ and $G$. Table \ref{tab:alphayp0} shows the values of the constants
as linear combinations of $\alpha$ and $p_{0}$.

\begin{table}[h]
\begin{center}
\begin{tabular}{|c|c|c|}
  \hline
  \textbf{Constant} & \textbf{$\alpha$} & \textbf{$p_{0}$} \\
  \hline
  $A$ & $0.09469$ & $0.0309$ \\
  $B$ & $682.718$ & $57.114$ \\
  $C$ & $1.745$ & $0.719$ \\
  $D$ & $2.212$ & $0.752$ \\
  $F$ & $2.408$ & $0.998$ \\
  $G$ & $2.0104$ & $0.333$ \\
  \hline
\end{tabular}
\caption{Values of the integration constants of the fields which are
linear combinations of the parameters $\alpha$ and $p_{0}$.}\label{tab:alphayp0}
\end{center}
\end{table}

The final expression for the fields can be found in the Appendix. In
Figure \ref{fig:todosloscampos}, the behavior of the field for
different allowed values\footnote{Allowed values
for $\alpha$ y $p_{0}$ are detailed in Section 4.} of $\alpha$ and $p_{0}$ are observed.
Figure \ref{fig:todosloscampos}(A) corresponds to $\alpha=1/\sqrt{20078}$ and
$p_{0}=-0.02164$, which is the value for $p_{0}$ when $\alpha$ has the
pointed value. Figure \ref{fig:todosloscampos}(B) shows the graphic for
the field, for a null value of $\alpha$, and $p_{0}=0.22$, which is the
upper limit for $p_{0}$ coming from Friedmann equation.  Finally,
Figure \ref{fig:todosloscampos}(C) shows the unique case in which the field
does not diverge at the origin, due to the fact that the value of $\alpha$
is such that the linear term in $a^{-2}$ in $\phi_{1}$ (see
eq.\eqref{eq:fi1pars} in the Appendix) is zero for all $p_{0}$.

\begin{figure}[!h]
\begin{center}
\includegraphics[width=1.1\linewidth]{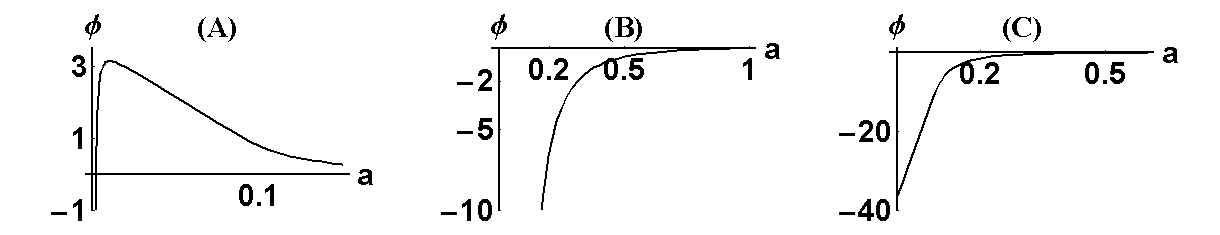}
\caption{Different behavior of $\phi$ vs $a$. (A) $\alpha=1/\sqrt{20078}$ and
$p_{0}=-0.02164$; (B) $\alpha=0$ y $p_{0}=0.22$; (C) $\alpha=-0.326062p_{0}$ and $p_{0}=0.22$.
In this case the field has no divergence at the origin, and its value there is -36.4087.} \label{fig:todosloscampos}
\end{center}
\end{figure}

\section{Experimental and observational data on $G(\phi)$}

\subsection{Corrections to Newtonian potential}

Theoretical speculations predict new effects
at distances of order less than 1mm. In particular, models
with spatial non-compact extra dimensions are of interest
because these "internal" dimensions could alter the form of the Newtonian potential.

If the extra dimension is non-compact, as in the case of RS model,
there is a continuous of Kaluza-Klein (KK) modes for the gravitational field.
The continuous spectra of KK modes leads to a correction to
the force between two static masses in the brane.
The potential for two point-like masses confined to the
brane reads \cite{RSalternative}
\begin{equation}\label{eq:potRS}
V_{\text{RS}}(r)=G\frac{m_{1}m_{2}}{r}\Big(1+\frac{1}{r^{2}k^{2}}\Big),
\end{equation}
where $k^{-1}$ should be of order of the distance of available
gravitational tests ($\sim$1mm ) or smaller.

Thus, in order to bound the deviation from the Newtonian potential,
one must constrain parameter $k$. Adelberger \emph{et al.}
\cite{hoyle-particle physics implications of a recent} performed experiments
with torsion balances to model the correction to Newtonian potential
using a power law of the form
\begin{equation}\label{eq:pothoyle}
\Delta V_{1 2}^{j}=-G
\frac{m_{1}m_{2}}{r}\beta_{j}\Big(\frac{1\text{mm}}{r}\Big)^{j-1},
\end{equation}
where the values of $j$ and $|\beta_{j}|$ are shown in Table \ref{tab:k}.

\begin{table}[h]
\begin{center}
\begin{tabular}{|c|c|}
  \hline
  $\textbf{k}$ & \textbf{$|\beta_{j}|(<)$} \\
  \hline
  $2$ & $4.5 × 10^{-4}$ \\
  $3$ & $1.3 × 10^{-4}$ \\
  $4$ & $4.9 × 10^{-5}$ \\
  $5$ & $1.5 × 10^{-5}$ \\
  \hline
\end{tabular}
\caption{Bounds on $|\beta_{j}|$ for $j=2,3,4,5$ obtained by
Adelberger \emph{et al.} \cite{hoyle-particle physics
implications of a recent}.}\label{tab:k}
\end{center}
\end{table}
For $j=3$ the extra term is
\begin{equation}\label{eq:deltaj3}
\Delta V_{1 2}^{3}=G
\frac{m_{1}m_{2}}{r}|\beta_{3}|\Big(\frac{1\text{mm}}{r}\Big)^{2},
\end{equation}
while the deviation predicted by RS model is
\begin{equation}\label{eq:deltaRS}
\Delta V_{1 2}^{\text{RS}}=G
\frac{m_{1}m_{2}}{r}\Big(\frac{1}{r^{2}k^{2}}\Big).
\end{equation}

Thus, the value of $k$ is constrained comparing
$\Delta V_{1 2}^{3}$ with $\Delta V_{1 2}^{\text{RS}}$.
Then, the characteristic scale at which the effects due to
the presence of an extra dimension become important is
\begin{equation}\label{eq:unosobrek}
\frac{1}{k} < 0.01\text{mm}.
\end{equation}

\subsection{Observational bounds on $G$ variation}

Bounds on the variation of the Newton constant are obtained from local and
cosmological observations.  Local observations are related to the solar
system, as well as nearby stars. Geological and paleontological data, as well
as planetary orbits, stellar densities, and luminosities, are of great
importance when studying $G$, because they are affected by its
variation.

\subsubsection{Bounds on $\Delta G/G$}

\emph{Planetary radius variation:}
In 1961, Egyed proposed that paleomagnetic data could
be used for the calculation of Earth paleoradius (past to current
planetary radius ratio) in different geological eras.
Starting from the hypothesis that the continental material area
remained constant during planetary expansion, Egyed found that
the ratio between current and past angular separation (paleolatitud)
of two given sites is proportional to the paleoradius \cite{egyed}.
A few years later, in 1978, McElhinny \emph{el al.} related Earth radius
variation to time evolution of gravitational constant,
and extended the analysis to the Moon, Mars and Mercury \cite{mcelh}.
According to their work,
\begin{equation}\label{eq:deltaRdeltaG}
\frac{\Delta R}{R}=-\gamma \frac{\Delta G}{G},
\end{equation}
where $\Delta R$ is the variation of the radius $R$ and $\gamma$
is a constant that depends on the planet structure.

On the other hand, there is another way to write this relation using the
paleoradius $R_{\text{a}}$:
\begin{equation}\label{eq:RadeltaG}
\frac{\Delta G}{G}=\frac{R_{\text{a}}-1}{\gamma}.
\end{equation}

Table \ref{tab:paleo} summarizes the results for the Earth, the Moon,
Mars\footnote{There are two different analysis for Mars: (A) assumes
a 19km expansion during the last 3600 million years; (B) supposes a 1km
variation on martian radius in the past 1000 million years. See \cite{mcelh} for details.} and Mercury.
\begin{table}[h]
\begin{center}
\begin{tabular}{|c|c|c|c|c|c|}
  \hline
  \textbf{Planet}& \textbf{$R_{\text{a}}$} & \textbf{$\gamma$} & \textbf{Time [$10^{9}$ years]} & \textbf{$|\Delta{G}/G|(<)$} \\
  \hline
  Earth & $1.020±0.028$ & $0.085±0.02$ & 0.4 & 0.62 \\
  Moon & $1.0000±0.0006$ & $0.0004±0.001$ & 3.9 & 1.5 \\
  Mars (A)& 0.9944 & $0.03±0.01$ & 3.6 & 0.12 \\
  Mars (B) & $1.0000±0.0003$ & $0.03±0.01$ & 1.0 & 0.01 \\
  Mercury & $1.0000±0.0004$ & $0.02±0.005$ & 3.5 & 0.02 \\
  \hline
\end{tabular}
\caption{Paleoradius ($R_{\text{a}}$) of the Earth, the Moon, Mars (A and B)
  and Mercury, and bounds on relative variation of $G$.}\label{tab:paleo}
\end{center}
\end{table}
These results assumes that the surface of each studied planet
acquired its current shape by the time indicated in the fourth column.

\emph{Big-Bang Nucleosynthesis:}
Bounds of a different sort come from cosmology.
In 1990, Accetta \emph{et al.} studied bounds on gravitational
constant value during primordial nucleosynthesis, considering
neutron mean life measurements \cite{accetta}. They determined D, $^{3}$He and $^{7}$Li
abundances while varying $G$, and how this variation affects barion to
photon ratio.
On the other hand, Copi \emph{et al.} recalculated relative variation of $G$
since BBN, but using only primordial D
abundance in quasars \cite{copi}.
In both works the constraint on relative variation of $G$ is
\begin{equation}\label{eq:cotadeltagnuc}
|\frac{\Delta G}{G}|_{\text{BBN}} < 0.4,
\end{equation}
which means that the relative variation of gravitational constant since
BBN is less than 40\%, at the 95\% confidence level\footnote{It is important to note that the constraint is for the absolute value of
$\frac{\Delta G}{G}|_{\text{BBN}}$.\\ In \cite{copi} the constraints are: $0.85<\frac{G_{\text{BBN}}}{G_{0}}<1.21$, at the 68.3\% confidence level, and
$0.71<\frac{G_{\text{BBN}}}{G_{0}}<1.43$, at the 95\% confidence level. $G_{0}$ is the present value of the Newton constant.\\
In this work we made use of the last constraint.}.

\emph{Cosmic background anisotropies:}
Power spectra of cosmic microwave background anisotropies (CMBA)
can be useful while constraining $G$ variations in cosmological scales.
In \cite{chan-chu}, gravitational constant stabilization (convergence to
its current value) and its relation to CMBA are studied in detail.
Two possible parametrizations of $G$ are considered: one, corresponding
to an instantaneous stabilization, and the other to a stabilization
linear with the scale factor $a$.

If the stabilization is linear, the relative variation of $G$ since
recombination ($z \sim 1000$) is
\begin{equation}\label{eq:cotadeltagcmb}
|\frac{\Delta G}{G}|_{\text{CMB}} < 0.1.
\end{equation}
at the 95\% confidence level\footnote{Again, we aware the reader that the constraint is for the absolute value of
$\frac{\Delta G}{G}|_{\text{CMB}}$.\\ In \cite{chan-chu} the constraints are: $0.95<\frac{G_{\text{CMB}}}{G_{0}}<1.05$, for a $G$ variation modeled
by a step function, and $0.89<\frac{G_{\text{CMB}}}{G_{0}}<1.13$, for a variation modeled by a linear function of the scale factor. Both constraints are
at the 95\% confidence level.\\
In this work we made use of the last constraint.}.
Thus, the relative variation of $G$ (its absolute value) since recombination is less than 10\%.

\subsubsection{Bounds on $\dot{G}/G$}

\emph{Lunar Laser Ranging} (LLR) has been measuring the position of
the Moon with respect to the Earth during more than thirty years, with
a precision of 1cm. The misions \emph{Appolo} 11, 14 y 15, and
russian-french \emph{Lunakhod} 1 y 4 carried retro-reflectors to the
Moon, which reflect laser pulses sent from the Earth.
LLR data are used to constrain Weak Equivalence Principle, post-newtonian
parameters and $\dot{G}/G$.

According to 2004 data in \cite{LLR}, the maximum variation allowed to
gravitational constant today is
\begin{equation}\label{eq:cotadotG}
\frac{\dot{G}}{G}(\tau_{0})=(4 ± 9)× 10^{-13}\text{yr}^{-1}.
\end{equation}

\subsubsection{Bound on $\ddot{G}/G$}

Now, let us discuss the constraints coming from observational bounds on
$\ddot{G}/G$.  For that purpose, we consider a model slightly different
to brane cosmology.

Other model which considers scalar fields is the
scalar-tensorial theory of Brans and Dicke of 1961.
This theory contains a scalar field governing $G$ dynamics.
In usual notation we have \cite{BD},
\begin{equation}\label{eq:BD}
G(\varphi)=\frac{1}{\varphi}.
\end{equation}

$G$ time dependence is described by
\begin{equation}\label{eq:Gdet}
G(\tau)\sim \tau^{-\text{n}},
\end{equation}
where $\text{n}=2/(4+3\omega)$, and $\omega$ is a model parameter\footnote{In Brans-Dicke
work $\omega=const.$, but there are more complex models where $\omega=\omega(\phi)$.},
which measures the deviation from General Relativity (GR).
GR results are reobtained when $\omega$ goes to infinity.
Considering \eqref{eq:Gdet}, it can be shown that
\begin{equation}\label{eq:ddotG}
\frac{\ddot{G}}{G}(\tau)=\text{n}(\text{n}+1)\tau^{-2}.
\end{equation}

Benvenuto, Althaus and Torres in \cite{BAT} gave a bound to the
absolute value of $\omega$, coming from white dwarfs evolution and its
relation with a varying $G$.  According to their results, the
calculated luminosities differ from the observed ones for
$|\omega|<5000$. Then, the allowed values range is $|\omega|>5000$.

Equation \eqref{eq:ddotG} evaluated today ($\tau_{0}$=13730 million years)
together with the constrain on $\omega$, give a bound on the variation
of the second time derivative of $G$:
\begin{equation}\label{eq:cotaddotG}
-3.55648 × 10^{-40} \text{s}^{-2} < \frac{\ddot{G}}{G}(\tau_{0})<
7.11082 × 10^{-40} \text{s}^{-2}.
\end{equation}

At this point, one could ask about the relation between Brans-Dicke
model and the one studied in this work. The explanation is the following: Brans-Dicke
theory in the limit $\omega\to\infty$ gives back GR; while GR and brane cosmology should
be equivalent in the studied limit.  Then, at first order, BD and
brane cosmology are equivalent and constraints on $\omega$ can be translated
into constrains on brane model parameters. However, it is not clear
that the differences between BD and brane cosmology lead to a great
discrepancy on the limit \eqref{eq:cotaddotG}. Then, we study the
inclusion and exclusion of this bound in following analysis.

\section{Constraining model parameters}

\subsection{Parameters $\alpha$ and $\dot{\phi}(\tau_{0})$}

$G$ variations, \emph{i.e.} $\Delta G/G$, $\dot{G}/G$ and
$\ddot{G}/G$, can be written in terms of $\phi$ as follows:
\begin{equation}\label{eq:deltaG1}
\frac{\Delta G}{G}(a)=-\alpha \phi(a),
\end{equation}

\begin{equation}\label{eq:gpunto1}
\frac{\dot{G}}{G}(a)= \alpha \dot{\phi}(a),
\end{equation}

\begin{equation}\label{eq:gdospuntos1}
\frac{\ddot{G}}{G}(a)=\alpha^{2} \dot{\phi}^{2}(a)+\alpha
\ddot{\phi}(a).
\end{equation}

Then, observational bounds discussed above restrict
the possible values of $\alpha$ and $p_{0}$.
The strongest restrictions on $\alpha$ and $p_{0}$ come from
BBN and recombination.
The largest value for the modulus of $\alpha$ is fixed
by the combination of BBN and CMB restrictions;
that is
\begin{equation}\label{eq:cotaalpha}
|\alpha|< \frac{1}{\sqrt{20078}}=0.007,
\end{equation}
while $|p_{0}|<0.22$.

\begin{figure}[h]
\begin{center}
\includegraphics[width=1.1\linewidth]{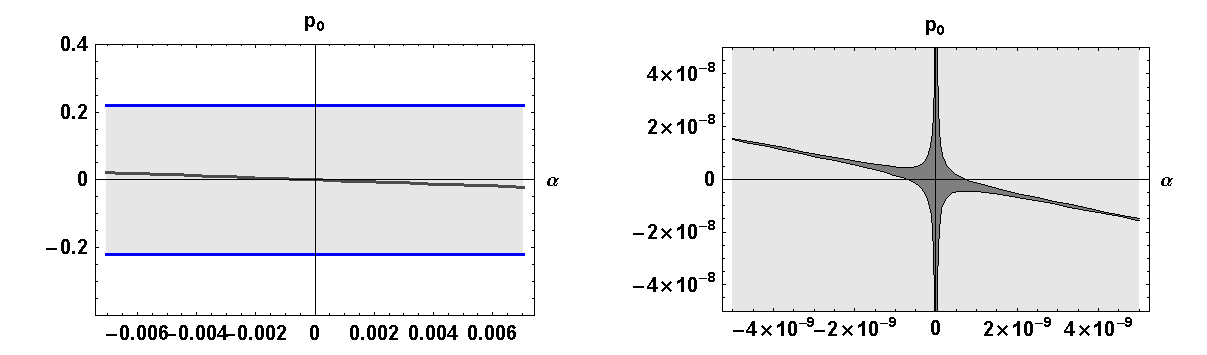}
\caption{Allowed values for $\alpha$ and $p_{0}$. The figure on the right is
a zoom-in of the left one. Allowed values are located inside the
dark gray contour, and satisfy $|\alpha|<0.007$ y $|p_{0}|<0.22$. Blue
lines in the left figure establish the bound for the largest value for $|p_{0}|$, a condition coming from
Friedmann equation today.}
\label{fig:cruzsolucion}
\end{center}
\end{figure}

The allowed values are shown in Figure \ref{fig:cruzsolucion}.
Allowed values are inside the surface which perimeter is
given by the green hyperboles.

In \cite{brax-davis-cosmological solutions of supergravity in singular
  spaces}, Brax and Davis find two theoretical values for $\alpha$, emerging
from supergravity in singular spaces. This values are $\alpha=1/\sqrt{3}$ and
$\alpha=-1/\sqrt{12}$.  From the analysis we have just done, based on observational
constraints, the absolute value of $\alpha$ is bounded by \eqref{eq:cotaalpha}
and then, it excludes these theoretical values.

\subsubsection{Statistical analysis}

Now, we would like to investigate how robust our constraints are.
That is, we will study how the constraints change if one excludes
ones or others data.

Excluding $G$ relative variation since Big-Bang nucleosynthesis data, and using the other
data, \emph{i.e.} $\Delta G/G|_{\text{CMB}}$, $\Delta G/G|_{\text{paleoradii}}$,
$\dot{G}/G|_{\text{today}}$, $\ddot{G}/G|_{\text{today}}$ and $|p_{0}|<0.22$, it can
be deduced that the strongest constraint on $\alpha$ absolute value is given by
the combination of $\Delta G/G|_{\text{CMB}}$ and $\ddot{G}/G|_{\text{today}}$.

In this case, we have
\begin{equation}\label{eq:alphasinbbn}
|\alpha|<\frac{1}{\sqrt{1787}}=0.024.
\end{equation}

If we exclude CMB data, in addition to BBN data, and make a similar
analysis, we have that the constraint on $|\alpha|$ is fixed by the combination
of $\ddot{G}/{G}|_{\text{today}}$ and $\dot{G}/G|_{\text{today}}$:
\begin{equation}\label{eq:alphasinbbn+cmb}
|\alpha|<\frac{1}{\sqrt{237}}=0.065.
\end{equation}

Omitting observational data of the second derivative of $G$, we obtain
the same results that those at the beginning of this subsection. This is because the
severest constraint on $|\alpha|$ is given by the combination of BBN and CMB
data.

If we exclude BBN data, in addition to $\ddot{G}/G|_{\text{today}}$, $\alpha$ is
constrained by the combination of $\Delta G/G|_{\text{CMB}}$ and
$\dot{G}/G|_{\text{today}}$ cotes, being
\begin{equation}\label{eq:alphasinbbn+ddotg}
|\alpha|<\frac{1}{\sqrt{243}}=0.064.
\end{equation}

Finally, if we only take into account the bounds coming from today´s
value of $G$ derivative, and those related to planetary paleoradius
(\emph{i.e.} BBN, CMB and $\ddot{G}/G|_{\text{today}}$ limits, excluded), $|\alpha|$ has no
bound.

Table \ref{tab:alpha1} condenses the results obtained in data analysis.

\begin{table}[h]
\begin{center}
\begin{tabular}{|c|c|}
  \hline
  $|\alpha|$ & \textbf{Obtained with:} \\
  \hline
  $1/\sqrt{3}=0.577$& SUGRA \\
  $1/\sqrt{12}=0.289$ & SUGRA \\
  $1/\sqrt{237}=0.065$ & $\ddot{G}/G|_{\text{today}}$; $\dot{G}/G|_{\text{today}}$  \\
  $1/\sqrt{243}=0.064$ & CMB; $\dot{G}/G|_{\text{today}}$ \\
  $1/\sqrt{1787}=0.024$ & CMB; $\ddot{G}/G|_{\text{today}}$ \\
  $1/\sqrt{20078}=0.007$ & BBN; CMB \\
  no limit & $\ddot{G}/G|_{\text{today}}$; Mercury paleoradius \\
  \hline
\end{tabular}
\caption{Constraints on $\alpha$ obtained combining $G$ variations
  observational data.  The first two values are equalities found with
  supergravity (SUGRA) in singular spaces. The other values are cotes and
  should be read ''$|\alpha|< ...$''. The severest constraint is given by
  Big-Bang nucleosynthesis (BBN) and CMB. All the constrains, except the last one, exclude the first
  two values.}\label{tab:alpha1}
\end{center}
\end{table}

In the studied cases, the upper bound on $|p_{0}|$ is fixed
by Friedmann equation evaluated today, \emph{i.e.} $|p_{0}|< 0.22$.
In addition, it should be said that $|\alpha|$ is always below
0.065, except in the last case, where its value has no bound.

Analog analysis can be done, but assuming a different bound
on BD parameter: $|\omega|<500$ (instead of $|\omega|<5000$).
Table \ref{tab:alpha2} shows the results.

\begin{table}[h]
\begin{center}
\begin{tabular}{|c|c|}
  \hline
  $|\alpha|$ & \textbf{Obtained with} \\
  \hline
  $1/\sqrt{231}=0.066$ & $\ddot{G}/G|_{\text{today}}$; $\dot{G}/G|_{\text{today}}$  \\
  $1/\sqrt{243}=0.064$ & CMB; $\dot{G}/G|_{\text{today}}$ \\
  $1/\sqrt{178}=0.075$ & CMB; $\ddot{G}/G|_{\text{today}}$ \\
  $1/\sqrt{20078}=0.007$ & BBN; CMB \\
  no limit & $\ddot{G}/G|_{\text{today}}$; Mercury paleoradius \\
  \hline
\end{tabular}
\caption{Constrains on $\alpha$ when combining observational
data of $G$ variation, in the case $|\omega|<500$.}\label{tab:alpha2}
\end{center}
\end{table}

The difference with respect to the results obtained with
$|\omega|<5000$ lies in the constraint of $|\alpha|$ found
combining CMB data and $\ddot{G}/G|_{\text{today}}$ bound.
In this case, the constraint on $|\alpha|$ is less restrictive
since it is three times bigger.

\subsection{5D parameters, $T$ and $\Lambda_{eff}$}

The 5-dimensional parameters of the model can also be constrained, as
well as the 4-D cosmological constant $\Lambda_{\text{eff}}$.  Using $H_{0}$ and
$\Omega_{\Lambda}$ from Subsection 3 of the Appendix, the constraint on $k$ from
\eqref{eq:unosobrek}, and $|\alpha|< 1/\sqrt{20078}$, we have the results
of Table \ref{tab:5dTL}.

\begin{table}[h]
\begin{center}
\begin{tabular}{|c|c|}
  \hline
  Parameter & Bound \\
  \hline
  $\kappa_{5}^{2}$ & $< 2.8 × 10^{-99}s^{3}$  \\
  $M_{5}=1/\kappa_{5}^{\frac{2}{3}}$ & $> 4.7 × 10^{8} GeV$ \\
  $\Lambda_{5}$ & $< -4.2× 10^{27} s^{-2}$ \\
  $|T|$ & $< 1 + 3.05× 10^{-63}$ \\
  $\Lambda_{\text{eff}}$ & $= 4.8 × 10^{-18} s^{-2}$ \\
  \hline
\end{tabular}
\caption{Constrains on 5D parameters, $T$ and $\Lambda_{\text{eff}}$.}
\label{tab:5dTL}
\end{center}
\end{table}

The same results are obtained when the other constraints on $\alpha$ from Tables
\ref{tab:alpha1} and \ref{tab:alpha2} are used.
Constraints on $\kappa_{5}^{2}$ and $M_{5}$ are in accord with
the ones predicted in \cite{lang2,maartens1,lang3}.

\section{Conclusions}

In this work, we studied a brane-world cosmological
model in which variation of the Newton coupling $G$ emerges
naturally. This model is inspired in supergravity in singular spaces
\cite{brax-davis-cosmological solutions of supergravity in singular spaces}.
By resorting to available observational data, we manage to constrain
the parameters of the theory.
Light elements abundances, coming from Big Bang nucleosynthesis (He, Li y
D)(eq. \eqref{eq:cotadeltagnuc}), CMB (eq. \eqref{eq:cotadeltagcmb})
and, near in time, planetary radii variations, allowed to constrain
the relative variation of $G$, \emph{i.e.} $\Delta G/G$ (see Table
\ref{tab:paleo}). Measurements of Lunar position with respect to the
Earth with LLR offered us a bound on today's value of the $G$ time
derivative to $G$ ratio (eq. \eqref{eq:cotadotG}).
Combining the severest constraints on $G$ variations, \emph{i.e.} those
coming from Big Bang nucleosynthesis and CMB, a bound for the absolute value of the parameter
$\alpha$ was obtained. In fact, this parameter must be less than $0.007$, and the
value of $|p_{0}|$ is bounded while evaluating Friedmann equation today.

Statistical analysis was performed on the results to analyze how robust
the bounds are against the exclusion of particular set of data.
That is, we studied how the upper bound of $|\alpha|$
gets affected if one excludes different sets of data. Results are
shown in Tables \ref{tab:alpha1} and \ref{tab:alpha2}.  It is
worth mentioning that the upper bound for $|\alpha|$ is always
less than $1/\sqrt{178}\sim 0.07$ (except in the
case where Big Bang nucleosynthesis and CMB data are excluded,
which turn out to be the most important ones).

Our analysis presents a method to investigate the phenomenological viability
of models that, among other features, predict time variation of the
fundamental couplings. It could be interesting to extend our analysis
to other brane-world type scenarios.

\subsection*{Acknowledgments}
The authors are grateful to G. Giribet for useful discussions. The work
of L.A. was supported by CONICET.

\appendix

\section{Field equations for the scalar field}

\subsection{Matter and radiation regime}

During matter and radiation epoch. Klein-Gordon equation reads

\begin{equation}\label{eq:KGMR}
\ddot{\phi_{1}}+4H_{1}\dot{\phi_{1}}+2\alpha
H_{0}^{2}\frac{\Omega_{\text{M}}}{a_{1}^{3}}=-16 \alpha H_{0}^{2}
\Omega_{\Lambda},
\end{equation}
where we used \eqref{eq:identrhoomega} and \eqref{eq:identlam}.

Differentiating with respect to $a_{1}$, we have

\begin{equation}\label{eq:psi1}
\frac{d\psi_{1}}{da_{1}}+\frac{4}{a_{1}}\psi_{1}=\frac{-2\alpha
H_{0}}{\sqrt{\Omega_{\text{M}}a_{1}+\Omega_{\text{R}}}}\Big(\frac{\Omega_{\text{M}}}{a_{1}^{2}}+8\Omega_{\Lambda}a_{1}\Big),
\end{equation}
where we used \eqref{eq:frMRstan1} and $\psi_{1}=\dot{\phi_{1}}$.

Being reminded of the fact that during this regime $a_{1}$ is small, the function on
the right hand side can be approximated by its first term in power expansion
for $a_{1}\sim 0$, thus

\begin{equation}\label{eq:psi1aprox}
\frac{d\psi_{1}}{da_{1}}+\frac{4}{a_{1}}\psi_{1}=-2\alpha
H_{0}\frac{\Omega_{\text{M}}}{\sqrt{\Omega_{\text{R}}}a_{1}^{2}}.
\end{equation}

Integrating, we find

\begin{equation}\label{eq:psi1dea}
\psi_{1}(a_{1})=\frac{-2H_{0}\Omega_{\text{M}}\alpha}{3\sqrt{\Omega_{\text{R}}}a_{1}}+\frac{A
H_{0}}{a_{1}^{4}},
\end{equation}
where $A$ is an integration constant.

To find $\phi_{1}(a_{1})$, it is necessary to integrate once again,
taking into account that

\begin{equation}\label{eq:dfida}
\frac{d\phi_{1}}{da_{1}}=\frac{1}{a_{1}H_{1}}\frac{d\phi_{1}}{d\tau}=\frac{1}{a_{1}H_{1}}\psi_{1}(a_{1}).
\end{equation}

Then, we have
\begin{equation}\label{eq:phi1int}
\phi_{1}(a_{1})=B+ \int da_{1}
\psi(a_{1})\frac{a_{1}}{H_{0}\sqrt{\Omega_{\text{M}}a_{1}+\Omega_{\text{R}}}}\simeq
B+ \int da_{1} \psi(a_{1})\frac{a_{1}}{H_{0}\sqrt{\Omega_{\text{R}}}},
\end{equation}
Where $B$ is a constant, and where we only considered the first term in the
power expansion of $\frac{a_{1}}{\sqrt{\Omega_{\text{M}}a_{1}+\Omega_{\text{R}}}}$.

The solution for the differential equation is

\begin{equation}\label{eq:fi1}
\phi_{1}(a_{1})=B-\frac{A}{2\sqrt{\Omega_{\text{R}}}a_{1}^{2}}-\frac{2\Omega_{\text{M}}\alpha}{3\Omega_{\text{R}}}a_{1}.
\end{equation}

\subsection{Matter and cosmological constant regime}

During this regime the Klein-Gordon equation is

\begin{equation}\label{eq:KGML}
\ddot{\phi}+4H_{2}\dot{\phi}+2\alpha
H_{0}^{2}\frac{\Omega_{\text{M}}}{a_{2}^{3}}=-16 \alpha H_{0}^{2},
\Omega_{\Lambda}
\end{equation}
which, written as a function of $a_{2}$ derivatives, is

\begin{equation}\label{eq:psi2}
\frac{d\psi}{da_{2}}+\frac{4}{a_{2}}\psi=\frac{-2\alpha
H_{0}}{a_{2}^{\frac{5}{2}}\sqrt{\Omega_{\text{M}}+\Omega_{\Lambda}a_{2}^{3}}}(\Omega_{\text{M}}+
8\Omega_{\Lambda}a_{2}^{2}).
\end{equation}

\begin{itemize}
\item First, consider $a_{2} \ll 1$:
\begin{equation}\label{eq:aprox1}
\frac{1}{a_{2}^{\frac{5}{2}}\sqrt{\Omega_{\text{M}}+\Omega_{\Lambda}a_{2}^{3}}}(\Omega_{\text{M}}+
8\Omega_{\Lambda}a_{2}^{2})\simeq
\frac{\sqrt{\Omega_{\text{M}}}}{a_{2}^{\frac{5}{2}}}.
\end{equation}
Replacing this equation in \eqref{eq:psi2}, we have

\begin{equation}\label{eq:psi2a2}
\psi_{2}(a_{2})=-\frac{4H_{0}\sqrt{\Omega_{\text{M}}}\alpha}{5
a_{2}^{\frac{3}{2}}}+\frac{C H_{0}}{a_{2}^{4}},
\end{equation}
where $C$ is an integration constant.

The solution for the field comes form the integral
\begin{equation}\label{eq:fi2aprox}
\phi_{2}(a_{2})=D+ \int da_{2}
\psi_{2}(a_{2})\frac{1}{a_{2}H_{2}}\simeq D+ \int da_{2}
\psi_{2}(a_{2})\frac{1}{H_{0}}\sqrt{\frac{a_{2}}{\Omega_{\text{M}}}},
\end{equation}
where $D$ is a constant, and \eqref{eq:frMLstan} and \eqref{eq:psi2a2}
were used. Also, $1/a_{2}H_{2}$ was approximated by its
first order in the power expansion, for $a_{2} \ll 1 $.
After integrating, we have

\begin{equation}\label{eq:fi2}
\phi_{2}(a_{2})=D-\frac{2C}{5\sqrt{\Omega_{\text{M}}}a_{2}^{\frac{5}{2}}}-\frac{4}{5}\alpha
\ln (a_{2}).
\end{equation}

\item Now, consider $a_{2} \sim 1$:
\begin{equation}\label{eq:aprox2}
\frac{1}{a_{2}^{\frac{5}{2}}\sqrt{\Omega_{\text{M}}+\Omega_{\Lambda}a_{2}^{3}}}(\Omega_{\text{M}}+
8\Omega_{\Lambda}a_{2}^{2})\simeq
\frac{8\Omega_{\Lambda}+\Omega_{\text{M}}}{\sqrt{\Omega_{\Lambda}+\Omega_{\text{M}}}}.
\end{equation}
As above, and making the approximation

\begin{equation}\label{eq:unosobreaH2}
\frac{1}{a_{2}H_{2}} \simeq
\frac{1}{H_{0}\sqrt{\Omega_{\Lambda}+\Omega_{\text{M}}}},
\end{equation}
we obtain the approximate value for the field

\begin{equation}\label{eq:fi3}
\phi_{3}(a_{2})=G-\frac{F}{3\sqrt{\Omega_{\Lambda}+\Omega_{\text{M}}}a_{2}^{3}}-
\frac{\alpha}{5(\Omega_{\Lambda}+\Omega_{\text{M}})}(8\Omega_{\Lambda}+\Omega_{\text{M}})a_{2}^{2},
\end{equation}
where $G$ and $F$ are constants.

\end{itemize}

\subsection{Final form for the field}

The values we take for the cosmological parameters appearing
in the expression of the fields
are $H_{0}=22.69 × 10^{-19}\text{s}^{-1}$; $\Omega_{\text{M}}=0.28$;
$\Omega_{\Lambda}=0.716$; $\Omega_{\text{R}}=4.6 × 10^{-5}$ (see
\cite{wmap} for further details).

The values for the integration constants can be found with the
parameters above and with the boundary conditions discussed in Section 2.
The value for the scale factor in the intermediate time $a_{\text{I}}$ is
$0.542$. Then,

\begin{equation}\label{eq:fi1pars}
\phi_{1}(a_{1})=682.718\alpha + 57.1142 p_{0}-\frac{6.9539\alpha +
2.2674 p_{0}}{a_{1}^{2}}-4029.68 \alpha a_{1},
\end{equation}

\begin{equation*}
\phi_{2}(a_{2})=-(543.028\alpha + 224.698 p_{0})+ (15241.3 \alpha +
6283.55 p_{0}) (a_{2}-a_{\text{u}})-
\end{equation*}
\begin{equation}\label{eq:fi2pars}
-(297480 \alpha + 122591 p_{0}) (a_{2}-a_{\text{u}})^{2},
\end{equation}

\begin{equation}\label{eq:fi3pars}
\phi_{3}(a_{2})=p_{0}(a_{2}-a_{0})-(6.0313 \alpha +
2p_{0})(a_{2}-a_{0})^{2},
\end{equation}
with $a_{\text{u}}=0.0897$ and $a_{0}=1$.

Notice that $\phi_{2}$ series is around $a_{\text{u}}=0.0897$, while $\phi_{3}$ series
is around
$a_{0}=1$.

\end{document}